# Acoustic Methods for Evaluation of High Energy Explosions

## Yury I. Lobanovsky

*IRKUT Corporation*

*68, Leningradsky prospect, Moscow, 125315, Russia*

E-mail: streamphlow@gmail.com

**Abstract**

Two independent acoustic methods were used to verify the results of earlier explosion energy calculations of Chelyabinsk meteoroid. They are: estimations through a path length of infrasound wave and through maximum concentration of the wave energy. The energy of this explosion turned out the same as in previous calculations, and it is close to 57 Mt of TNT. The first method, as well as evaluations through seismic signals and barograms, have confirmed the energy of Tunguska meteoroid explosion at 14.0 – 14.5 Mt level. Moreover, there is a good agreement between acoustic estimations and other data for the explosion energy of another meteoroid that was ended its flight over the southern part of Indian Ocean, and for two catastrophic volcanoes explosions – Bezymyanny and Krakatoa.

## I. Introduction

Explosion energy of blasts that occurred as the result of braking in the Earth's atmosphere of Chelyabinsk and Tunguska meteoroids was calculated on the basis of regular physical and mathematical procedures [1 – 3].There was shown by analysis of their characteristics that they belong to one and the same family of cometary debris [4]. These calculations were carried out in March 2013. Once were obtained detailed information concerning the incident Chelyabinsk, these calculations were performed again [5]. They have shown that Tunguska meteoroid exploded at the altitude of about 8.25 km, and the energy of its explosion was 14.5 Mt (megatons of TNT). The explosion of Chelyabinsk meteoroid was 4 times more powerful – 57 Mt, but due to the fact that it occurred at the height of almost 3.4 times greater – about 28.2 km, its effect on the underlying surface within the radius of 50 km was less strong than for Tunguska explosion. In this case, the closer to the epicenter of the explosion, the relatively weaker was the effect of more powerful, but much more high-rise Chelyabinsk explosion than Tunguska blast. At the epicenter overpressure peak on the shock wave from the explosion of Chelyabinsk meteoroid was 7 times lower than that of Tunguska, and therefore there were no serious damages from Chelyabinsk explosion on the ground. The most significant damage from it (not counting destruction sometimes of window and door structures and different gates) was that about 0.2 $km^2$ window glasses were broken up [6].

All these results are direct consequences of the basic laws of celestial bodies' motion, their explosive disintegration during entry to the Earth's atmosphere and of spread of shock waves, as well as data relating to the window glasses that was broken up in winter of 2013 and to tree-felling in summer of 1908. These patterns are based on conservation laws and, in principle, are quite simple. So the results presented in papers [1 – 3] were rather obvious for their author. And he was surprised that there was somebody to whom these results seemed to be a shock of fundamentals. But these «fundamentals» were formed by media leaning on the hasty judgments of some scientists, who, as believed, could estimate this phenomenon. However they have become hostage to a completely new situation, when the scale of the phenomenon was a lot more than that to which they previously accustomed to.

In this regard, it seems appropriate to make some estimations, which are clear and simple for every. In the first place, these estimations should be made for Chelyabinsk explosion energy, and such approaches should be different from those used in [1 – 5]. At present, this has become possible due to the advent of the Internet information about the acoustic characteristics of the Chelyabinsk explosion [7, 8]. This data are not full, but it was enough for estimations.

## II. Estimation of the explosion energy through a path length of infrasound wave

First, we consider the information about the path length in the atmosphere of sound wave caused by Chelyabinsk explosion. Its frequency was very low – about 0.03 Hz, so it is not perceived by ear, but it was registered with the aid of global network of infrasound stations. This wave was registered three times during its propagation in the Earth's atmosphere in Greenland [8] at a distance of 5.0 megameters (thousands of kilometers) from the epicenter of the explosion at I18DK station of International Monitoring System (see Fig. 1) located in the vicinity of the U.S. airbase Thule. Thus, the path length of the wave was at least 85 megameters [7].

It is known that heights of infrasound waves' channel reach up to about 110 kilometers at large distances [9, 10]. This quantity is negligible in comparison with the horizontal dimensions of the atmosphere, constituting at least 20 megameters – half the circumference of the Earth. Therefore, unlike the case of short distances from the blast, when the radius R, at which a predetermined overpressure on the shock wave may be reached, is described by formula (1)

$$R \sim E_e^{\frac{1}{3}}, \qquad (1)$$

where $E_e$ is the energy of the explosion [11], at such great distances two-dimensional waves' propagation process should be write down as

$$L \sim E_e^{\frac{1}{2}}, \qquad (2)$$

wherein the symbol L is using for the wave path length, as in this case it loses the character of the radius from the center of explosion (or epicenter, which are not distinguished in this scale).

From the formula (2) it follows that, other things being equal, the maximum path length of the wave sound $L_{max}$ to a first approximation is proportional to the energy of the explosion in the degree ½. Then, having the appropriate characteristics of a «baseline» explosion with energy $E^*$ and a maximum path length of infrasound wave $L^*$, the energy of any other explosions may be evaluated through the maximum path length of their sound waves:

$$E_a = E^* \left( \frac{L_{max}}{L^*} \right)^2 \qquad (3)$$

Here $E_a$ is the explosion energy from the acoustic estimates.

When comparing very powerful explosions of different physical nature there can be only two potential problems: their different energy spectra, and the impact of different heights of the explosion. It is known that on the primary and secondary ionizing radiation of nuclear explosions, which is absent in other types of explosions considered here, is spent up to 15 % of its energy, while on the primary radiation – only one-third part of this value [12]. Usually, however, the energy of the secondary radiation do not usually counted in the TNT equivalent of a nuclear warhead, and there is no secondary ionizing radiation in a purely thermonuclear explosion. One of the basic points of reference was here the most powerful man-made thermonuclear explosion of so-called Tsar Bomb (AN-602), which was produced by the Soviet Union 30 October, 1961 at Sukhoy Nos test site on Novaya Zemlya [13]. Nuclear «ignition» energy was only 2.5 % of full energy of thermonuclear warhead, so that the energy of the secondary radiation from this explosion was very small and the remaining share of the primary ionizing radiation in 5 % may be neglected in these estimations also.

In addition, a possible reduction in fraction of light radiation's energy and, therefore, increasing the proportion of energy of a shock wave in the thermal explosion type of Chelyabinsk compared to the thermonuclear explosion, due to the lower temperature of the fireball at the initial stage of its development, may partly offset the impact of the explosion height. The higher the altitude, the less the density of the air, and the smaller part of energy of the shock wave is contained in total energy of explosion. Celestial bodies explode usually higher than thermonuclear warheads during their atmospheric tests, excluding the relatively small number of high-rise and cosmic blasts. At the height of 28.2 km as at the explosion of Chelyabinsk meteoroid the density of air is in 33.5 times lower than at the height of Tsar Bomb explosion, that was about 4.2 km [13, 14]. So there is every reason to expect that for Novaya Zemlya reference point in the formula (3), in which there is no influence of height, the energy of Chelyabinsk explosion should be lower than really.

It has been reported that the infrasound wave from Novaya Zemlya skirted the Earth three times [15]. In reality this means that there were such infrasound stations, which were recording it three times (and, they had to register two times the wave coming from the opposite direction). The distance between Sukhoy Nos test site on Novaya Zemlya and Chatham Island, where New Zealand I36NZ infrasound station of International Monitoring System (ISM) is located [16], where, according to the source [17], this wave was recorded for the last time, is 15.8 megameters and measured length of this wave path was not less than 95.8 megameters.

To account for the influence of the height of the explosion, we may take another point of reference, which is the most powerful explosion from U.S. Air Force series of acoustic observations of bolides, conducted in 1960 – 1974 years. This explosion with the energy of about 1.1 megatons occurred March 8, 1963 on the south-west of Indian Ocean, near its boundary with Atlantic Ocean at 51º south latitude and 24º east longitude in about 1800 km south of

Cape Agulhas, which is the southernmost point of Africa. According to the source [18] infrasound wave was recorded in Azores at the distance of 11.3 megameters from the epicenter of the explosion (I42PT infrasound station of ISM). Its height was not reported, but U.S. Air Force release informed in that «usually meteoroids disintegrate at altitudes of 30 to 45 km above the Earth's surface, but some penetrate the atmosphere to altitudes of about 20 km» [19]. So this Southern Bolide blew up certainly much higher than the Tsar Bomb, and since it was very big, it can be expected that the height of its blast was close enough to the bottom of the range specified in this release.

All these data and the values of the explosion energy are shown in Table 1, where $E_e$ is the explosion energy in megatons, $E_a$ is its score through a maximum length of the wave path $L_{max}$ by formula (3), measured in megameters. By definition, for the base explosion is $E_a = E_e = E^*$. The energy of Novaya Zemlya explosion was taken from the source [13], the energy of Chelyabinsk explosion – from sources [5].

**Table 1**

| N | Explosion | Year | H (km) | $L_{max}$ (Mm) | $E_a$ (Mt) | $E_e$ (Mt) |
|---|---|---|---|---|---|---|
| 1 | Novaya Zemlya | 1961 | 4.2 | 95.8 | 58 | 58 |
| 2 | Southern Ocean | 1963 | – | 11.3 | 0.81 | 1.1 |
| 3 | Chelyabinsk-1 | 2013 | 28.2 | 85 | 46 | 57 |
| 4 | Chelyabinsk-2 | 2013 | 28.2 | 85 | 62 | 57 |
| 5 | Chelyabinsk-3 | 2013 | 28.2 | 85 | 54 | 57 |

The deviation of the explosion energy of Southern Bolide from the acoustic estimation derived with formula (3) through Tsar Bomb explosion energy was about –26 % that also demonstrates considerable height of this bolide explosion. Table 1 shows that using of formula (3) for estimation of Chelyabinsk explosion energy through the low-altitude blast of Tsar Bomb (version Chelyabinsk-1) leads to the value of 46 Mt (deviation is –19 %), but high-rise explosion in Southern Ocean leads to the 62 Mt (version Chelyabinsk-2, deviation is +9 %). The average of these two estimations (Chelyabinsk-3) is 54 Mt, which is only on 3 Mt (5 %) lower than value obtained by calculations in paper [5].

Thus, we can believe, that taking into account the influence of altitude, acoustic method of explosion energy assessment may give quite acceptable accuracy. Since the effect of altitude is though quite noticeable, but not too strong, it will be considered using a simple linear approximation. Then formula (3) can be transformed as follows:

$$E_a = E^* \left( \frac{L_{max}}{L^* - kH} \right)^2, \qquad (4)$$

where H is the height of the explosion, and k is the influence coefficient of the explosion height on the maximum path length of infrasound wave. Thus, we can obtain the energy of the explosion regardless of its nature from formula (4) for a given characteristic values $E^*$, $L^*$ and k for maximum path length of infrasound wave.

Three more explosions are added to three that were considered earlier: explosions of Tunguska meteoroid [5] and volcanoes Bezymyanny and Krakatoa. Shock wave from Tunguska explosion was recorded twice in London [20, 21], which is remote from the epicenter of almost on 6 megameters as shown in line 4 of Table 2. In the middle of the XX century was another very significant incident with known parameters – a catastrophic explosion («lateral blast») of volcano Bezymyanny on Kamchatka [22]. Its energy is estimated to be $5 \cdot 10^{16}$ joules [23], which is approximately equal to 12 Mt. Maximum length of the wave path of this natural disaster has been specified in reference [21].

**Table 2**

| N | Explosion | Year | H (km) | $L_{max}$ (Mm) | $E_a$ (Mt) | $E_e$ (MMt) |
|---|---|---|---|---|---|---|
| 1 | Novaya Zemlya | 1961 | 4.2 | 95.8 | 59.7 | 58.0 |
| 2 | Chelyabinsk | 2013 | 28.2 | 85 | 56.8 | 56.8 |
| 3 | Southern Ocean | 1963 | 34.9 | 11.5 | 1.1 | 1.1 |
| 4 | Tunguska | 1908 | 8.25 | 46 | 14.2 | 14.4 |
| 5 | Kamchatka | 1956 | 3.1 | 42.5 | 11.7 | 12 |
| 6 | Sunda | 1883 | 1.5 | 415 | 1100 | 1090 |

Using the formula (4) and the least squares method, we define the parameters $E^*$, $L^*$ and k, which leads to the smallest discrepancy between the values of energies $E_a$ and $E_e$ considered in all points except the blast in Southern

Ocean, since its height was unknown. Then the values of the basic parameters in the formula (4) were as follows: the characteristic value of the explosion energy $E^* = 59.6$ Mt, path length waves at sea level $L^* = 97.2$ Mm and a height of influence coefficient k = 360. If the height is measured in kilometers (km), and running sound wave megameters (Mm), k = 0.36. The heights and energies of explosions in all six cases are presented in Table 2. Standard deviation of acoustic estimates of energy blasts from the original data does not exceed 2 %.

Based on these data, we can conclude that the acoustic energy estimations of low-height Tunguska and Kamchatka thermal explosions by formula (4) are in excellent agreement with the calculated data previously available. The estimation of Tunguska explosion energy from seismograms leads to the value of its energy 12.5 ± 2.5 Mt, and from barograms – to 12 ± 2.5 Mt [24], which also is in good agreement with those given in Table 2 values. The same can be said about the height of the explosion – 8.5 km [25] from the same source is different from the calculated value of 8.25 km (see Table 2) [5] on 0.25 km only.

In the case of the explosion of Southern Bolide from the equation $E_a = E_e$ it was calculated the estimated height of explosion. Since there are no other infrasound stations in the vicinity of Azores, the path length from infrasound waves of explosion could well be much larger than the range of its actual registration – 11.3 Mm. And this kind of errors because of the relative rarity of a network of infrasound stations is particularly large for such relatively short distances. Because of this the path length wave was rounded up to 11.5 Mm. The height of the explosion turned approximately equal to about 35 km. That is bound on the height of Southern Bolide explosion close enough to the bottom of the range of bolides' typical explosion heights according to the U.S. Air Force, which, apparently, was determined on the basis of a series of observations, including this event. By increasing the path length of 0.1 Mm assessment height of the explosion of Southern Bolide is reduced by 1.9 km. So in reality, this height could well be somewhat lower.

If for the most likely entry velocity of the celestial bodies into the atmosphere 17 km/s in accordance with calculation results of module of air explosions [26], the blast energy of 1.1 Mt at a height 35 km should be achieved for chondrites mini asteroid with density of 3000 kg/m$^3$, diameter of 39 m and mass of 92 kt. Its entry angle into the atmosphere had to be about 6.9º, almost like Chelyabinsk meteoroid (7.2º, see [5]). If it was comet fragment, then its density should be of about 570 kg/m$^3$, its diameter would be equal to 57 m and mass was 55 kt for the entry angle of about 15º. So the approximate evaluations of characteristics of the Southern Bolide and the parameters of its explosion seem to be quite adequate, and its main parameters are shown in Table 2. In this case, if the actual height of the explosion was lower, the angle of entry of the Southern Bolide was greater.

To estimate data about the energy of the most powerful among the considered here explosions – blast of Krakatoa, which occurred August 27, 1883 in Sunda Strait between the Indonesian islands Java and Sumatra [27, 28], is required some more details. And they have been considered in annex of this paper. Determinations of wave's length path and height of the explosion were simple. Now the height of volcano Rakata, which has remained from Krakatoa, is a little more than 0.8 km [27, 28 ], and the height of the mountain before the disaster was about twice as high than it is now [29], so the height of the explosion in 1883 was about 1.5 km.

The infrasound wave of this explosion went around the Earth according to different data from 7 to 11 times [27, 28]. In the last quarter of the XIX century accurate pressure measurements could be carried out in the area from St. Petersburg (Russia) to Boston (USA), which are removed from the epicenter of Sunda explosion at distance of 10 – 16 Mm (Pulkovo observatory and the Massachusetts Institute of Technology were already 2 – 4 decades at that time). Thus, the infrasound waves have run of order of 15 Mm before eleventh registration, and earlier ran over the Earth of 10 more times. All this lasted for more than 15 days, and may not always have the patience to spend a full cycle of measurements, and therefore in some places have been obtained smaller numbers of this value. Therefore, the maximum number of barograph registrations of this wave was 11 and it was used to estimate path length of wave, which was not less than 415 Mm (see Table 2). The calculated energy of the explosion of Krakatoa was equal to $E_e = 1.09 \pm 0.05$ Gt (gigatons), and almost coincided with her acoustic evaluation of $E_a = 1.10$ Gt, see line 6 of Table 2.

We now estimate the accuracy of determination of maximum path length of infrasound wave. Network of infrasound stations of International Monitoring System (IMS) under the Comprehensive Test Ban Treaty (CTBT) is shown in Fig. 1 [16]. Earlier this network had 60 stations [30] and by 2011, as follows from this figure 1, are remaining 41. With such number of them, the average distance between them is 3.8 megameters. However, there is still a considerable number of more or less similar infrasound stations not included in the IMS network. If all of these stations are of the order of 100, the average distance between them is equal to 2.4 megameters. It follows that the average error in determining the maximum path length of infrasound waves, which is equal to a half of these distances, may be, according to these data, about 1 – 2 megameters, that at explosion energies of 100 – 10 Mt leads to error in their determination about 2 – 10 %. According to Table 2, these differences do not exceed 2 %.

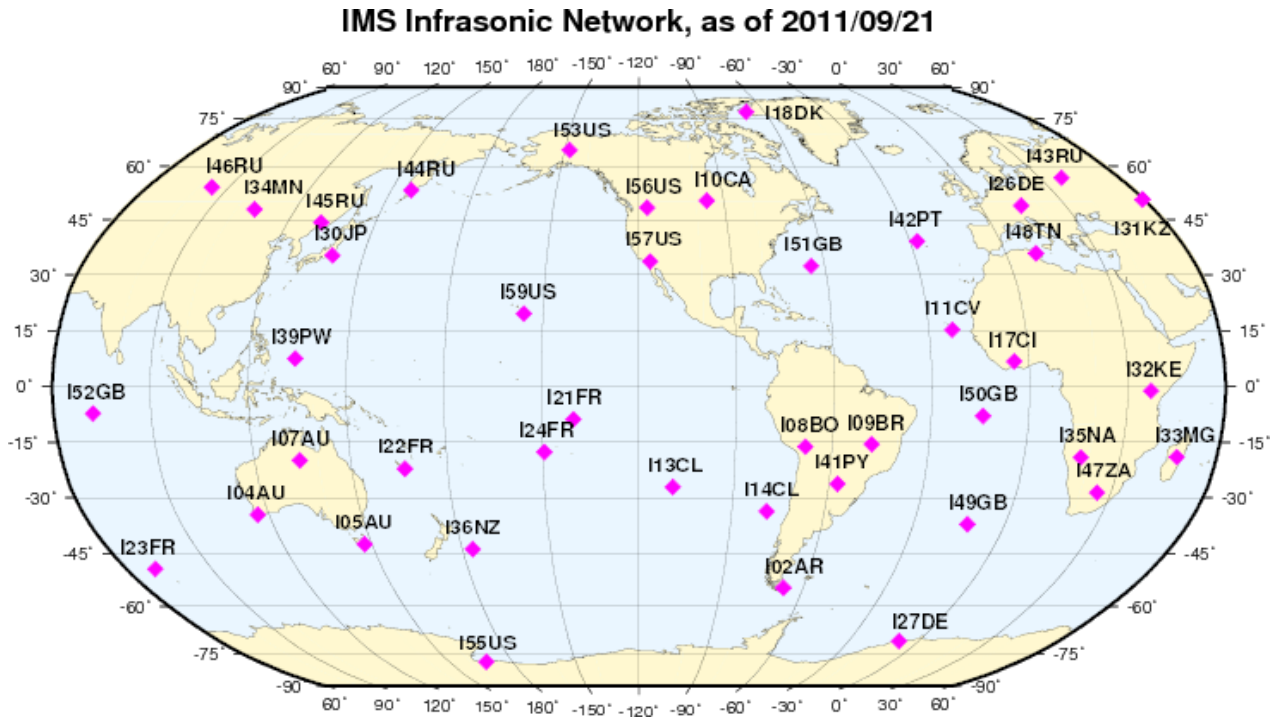

Fig. 1

Thus, the energy of Chelyabinsk meteoroid explosion which is 57 Mt, supported by data from infrasound stations, and the formula (4) can be used to estimate the energy of the powerful (more than 1 Mt) explosions up to gigatons level of energy in the Earth's atmosphere at the altitude of at least 35 – 40 km.

**III. Assessing the impact of the explosion energy to the frequency of its maximum concentration**

We will estimate now the energy of Chelyabinsk explosion by another acoustic method. It was learned that the acoustic «energy of the Chelyabinsk blast was concentrated at 0.03 hertz» [8]. Older and much more detailed data from the source [31] show that this frequency was equal to $0.030 \pm 0.007$ Hz. It is obvious that the source of the shock wave and then the acoustic wave is the so-called fireball. Without going into the details of the process we can simply considered it as a half-wave dipole antenna. Then the length of the emitted wave will be equal to twice its size. The maximum radius of Tsar Bomb fireball $R_{max}$ was 4.6 km [13]. The wave speed c at this point is independent of the explosion energy and was of about 0.54 km/s [32], since the only change is the distance at which it is achieved. Then the wave frequency $\nu^*$ is equal to:

$$\nu^* = \frac{c}{4R_{max}} \qquad (5)$$

Substituting into formula (5) numerical values of Novaya Zemlya explosion, we find that $\nu^* = 0.029$ hertz, which is different from the frequency of maximum energy in Chelyabinsk explosion only on 3 %. Thus, this agreement of calculated and experimental frequencies confirms once again the approximate equality of the energies of Chelyabinsk and Novaya Zemlya explosions.

Since as any size, which is associated with an explosion, can be recalculated in this scale with the aid of formula (1), at least in a first approximation, the velocity of the shock wave at the moment of stopping the growth of the fireball does not depend on the energy of the explosion, and according to the available experimental data, formula (5) may be converted into a form that is typical for empirical formulas relating to the explosions:

$$\nu^* = 0.116 \, E_e^{-\frac{1}{3}}, \qquad (6)$$

where the frequency $\nu^*$ is measured in hertz, and the explosion energy $E_e$ – in megatons. Apparently, the formula (6) can evaluate the frequency of maximum energy with an accuracy of 20 – 30%. It would be a good idea to verify this formula (6) by the results of all 10 powerful atmospheric thermonuclear blasts of energy more than 10 Mt, of which the Soviet Union had made 7 excluding Tsar Bomb explosion [33], and in the USA – 2 [34, 35]. More, there were two Soviet blasts with energy in the range of 5 – 10 Mt [33] and one US [36], which data won't be less useful to check. This information is probably stored still in some archives.

**IV. Annex: Determination of Sunda explosion energy**

Multiplicity recording the acoustic signal from the explosion of Krakatoa volcano in Sunda Strait of 7 – 11 times is quite confidently stated in the known sources of information [27, 28]. But there are huge differences in the estimations of the energy of this explosion. Wikipedia, which is usually quoted the most widespread opinions, said that the explosion energy of volcano Krakatoa was, ostensibly, 200 Mt (0.2 Gt) of TNT, and it is in 4 times more than Tsar Bomb explosion on Novaya Zemlya [27]. However, another encyclopedic resource argues that the energy explosion of Krakatoa was 1.5 Gt of TNT [37]. At the same time, a book published by the American Geological Society says that «the energy of the largest natural explosion of historical time – the explosion of Krakatoa in 1883, is estimated to be $10^{24}$ erg» [38], that is $10^{17}$ joules, or 24 Mt, which is 2.5 times less than Tsar Bomb explosion!

And at the same time is almost universally accepted that the explosion of volcano Tambora April 10, 1815 [39] was about 4 times more powerful than «the largest natural explosion of historical time» in Sunda Strait. Energy estimations for this blast, which occurred in early XIX century, are 20 Gt [37] and 24 Gt [40] that should lead to the values of energy in the 5 – 6 Gt for Krakatoa explosion. Moreover, in a documentary film about the catastrophic eruption of volcano Tambora «The Year without Summer» one respected volcanologist said that «... the explosion power in 3 million times exceeded the explosion power of Hiroshima» [41], that is about 45 Gt. Thus, in principle, the energy estimate of in Krakatoa explosion should be doubled up to 10 – 12 Gt. So, it is clear that volcanologists over the past 130 years have not been possible to determine this energy, since their extreme estimates differ by almost 500 times! These differences appear to be excessive, even for such very inexact science as volcanology, and even 120-fold error of Chelyabinsk explosion energy (see [1 – 3, 5]) does not seem too much in comparison with these results. For this reason the energy of Krakatoa explosion was calculated independently by the author of this work. Description of this calculation process is given below.

As mentioned previously, physical nature of explosion plays a minor role in the formation and development of shock waves. If the conditions, under which the explosions occur, are the same, the shock waves in the atmosphere should be the same also regardless of whether is this thermonuclear explosion or entry into the atmosphere of a celestial body or volcano explosive eruption. Of course, in a small neighborhood of volcano, the conditions, under which the explosion occurs, are very different from those that take place in the atmospheric nuclear blast or destruction of the meteoroid, since the volcano itself forms other boundary conditions as a huge mountain. However, at distances much greater than characteristic size of volcano, these differences are leveled, as appears from any wave theory. In addition, the explosion in the atmosphere of fast moving meteoroid occurs at almost its full stop, and, if we exclude the impact of ballistic shock wave, this blast is close to explosion of thermonuclear warhead with the same energy (see [1 – 3, 5]).

An additional argument that «all cats are gray at night» is in comparing of the first 5 explosions discussed in Table 2 of this article: regardless of their physical nature in all cases there is a good agreement of energies and their assessments through the characteristics of the sound wave. But a sound wave is a limiting case of a shock wave. In addition, the simplified «quasi-static» consideration of the explosive disintegration of the celestial bodies in module for calculating the consequences of a meteoroid impacts on the Earth [26] in this situation is not a disadvantage, but dignity, allowing estimating the effects on the underlying surface of any explosion in the air from meteoroids right down to the volcanoes. It is necessary to calculate only the entering into the atmosphere of «virtual meteoroid», decay of which leads to the explosion with a given energy at a given height.

In modeling the effects of various explosions we can choose any suitable «virtual meteoroid» with appropriate entry conditions to get right height and energy of the explosion, and it is a simple consequence of the structure of this computational module [26]. Thus, this module with using of «virtual meteoroids» may be used to estimate the explosion influence of any kind on the underlying surface, provided that the distance from the center of this blast to investigated point is much greater than the size of the exploding object.

In this way, we have calculated the energy of Sunda explosion, that is, volcano Krakatoa catastrophic eruption, which has occurred 130 years ago, in August 1883. Data about window glasses, broken in Batavia (Jakarta) [28, 42] and tree-felling in the equatorial forests along the banks of the strait [28] give the boundary conditions that determine the energy of the explosion at given height of 1.5 km. It was considered a lot of options, and the results of five of them, which are the most representative, is possible to see in Table 3. Here p is the overpressure peak on the shock wave in kilopascals at a distance L from the explosion, measured in kilometers along the ground and demonstrated in the column to the left of the pressure, $\xi$ is the number of registrations of infrasound wave coming by the shortest path from explosion.

**Table 3**

| N | Explosion | $L_1$ (km) | $p_1$ (kPa) | $L_2$ (km) | $p_2$ (kPa) | $L_3$ (km) | $p_3$ (kPa) | $E_e$ (Mt) | $\xi$ |
|---|---|---|---|---|---|---|---|---|---|
| 1 | **Sunda-1** | 29.0 | 30.0 | 130 | 3.2 | 155 | 2.6 | 200 | 5 |
| 2 | **Sunda-2** | 36.5 | 30.0 | 130 | 4.2 | 155 | 3.4 | 400 | 7 |
| 3 | **Sunda-3** | 49.3 | 30.0 | 130 | 6.4 | 155 | 5.0 | 1040 | 11 |
| 4 | **Sunda-4** | 50.0 | 30.0 | 130 | 6.5 | 155 | 5.1 | 1090 | 11 |
| 5 | **Sunda-5** | 51.5 | 30.0 | 130 | 6.8 | 155 | 5.3 | 1200 | 12 |

Distance $L_1$ is the radius of tree-felling zone on flat open countryside. Despite the fact that there are only qualitative estimates of this zone size since known sources inform about «air waves» … «which brought down the equatorial forests on Sunda Strait coast» [28], the topography of Sunda Strait and position of Krakatoa Island transforms the qualitative statement in quantitative. Sunda Strait coasts are formed from north shores of Sumatra Island and south-east shores of Java Island (see Fig. 2). Minimum distance from the epicenter to the north-west coast of Sumatra is 41.0 km and to the north-east coast is 36.4 km, while to the east coast of Java is 47.2 km. Thus, to the shock wave of Krakatoa explosion could «catch» rain forests on the both sides of Sunda Strait with the overpressure peak on the shock wave of 30 kPa (see [1 – 3, 5]), it should be achieved at the distance of no less than 48 – 49 km from the epicenter of the explosion. Distance $L_1$ = 50 km is shown in Fig. 2 with the aid of red line. The epicenter of the explosion is the point at the base of mark 1. Not shown in this figure plots of this circle pass in the open ocean where there is no jungle.

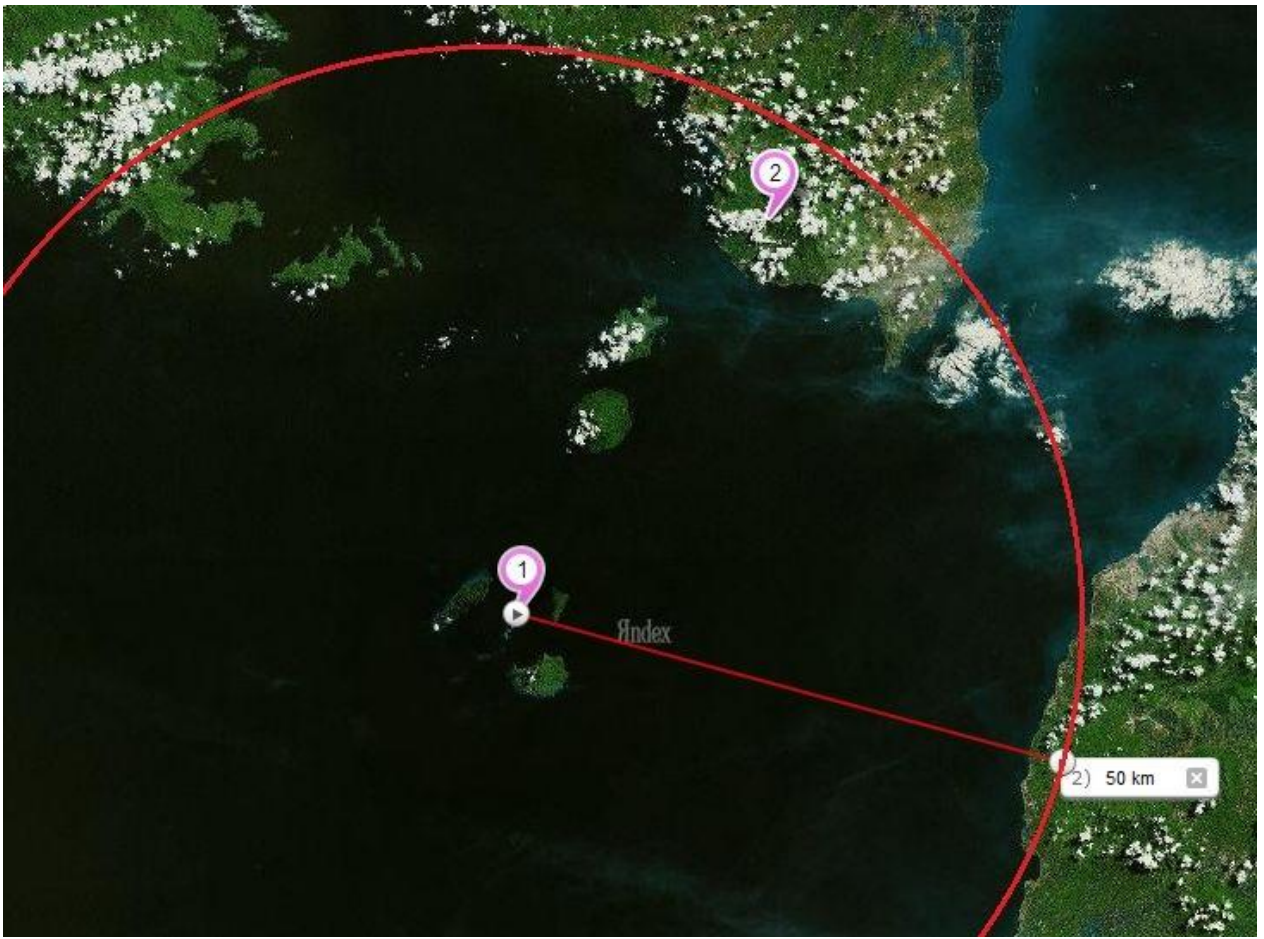

Fig. 2

In the Discovery Channel's documentary film about this event, as well as in publications based on report of R. Verbeek, which was the eyewitness of the disaster, was said: «The shock wave shattered windows in a radius of 130 km» [43], «eruption … caused windows to break at 150 km from source» [42], «air wave were shattered all the windows within a radius of 150 km» [44], «... the air wave tore roofs off houses and door off its hinges in Jakarta at a distance of 150 km from the crash point» [28]. The length $L_3$ = 155 km is the distance from the epicenter of the explosion to the main European district in Batavia (now Jakarta), the capital of the Dutch East Indies (Indonesia) of XIX century. This settlement was located near modern square of Jakarta, named Medan Merdeka [45, 46]. The length $L_2$ = 130 km is the distance to the suburban settlements around Batavia. Here there are located now the western satellite town of Jakarta Tangerang and the airport.

It is clear that the first three statements are more or less consistent with each other and with the actual topography of Batavia at that time, the fourth statement from Wikipedia is somewhat contradict them, exaggerating the destructions in the city (but the third statement describes also a somewhat more significant consequences in the center of Batavia at the time of Krakatoa eruption, compared with the first two). However the article in Russian Wikipedia is at best a tertiary review of secondary sources. And, at least, it is the double translation – from Dutch to English and from English to Russian, which leads often to a distortion of the interpretation or meaning of the message. Thus, the analysis of these data leads to the following picture of the devastation from the shock waves of Krakatoa explosion: there was broken noticeable part of window glasses in the European settlement of Batavia, at the distance of about 155 km from the epicenter, and in western townships at the distance of 130 km have been broken almost all glass windows, and may have corrupted some lightweight and weak doors and roofs, which are typical for tropical houses. This should mean that at distance $L_2$ overpressure peak on the shock wave was about 7 kPa and at distance $L_3$ – about 5 kPa. Together with the boundary condition for tree-felling at distance $L_1$ = 50 km (overpressure peak is 30 kPa), we have three boundary conditions to determine the only one unknown parameter – energy of Krakatoa explosion. This is so that a possible change in the height of the explosion at 100 – 200 m is almost negligible. Thus, in determining of explosion energy by using these three remaining boundary conditions, two of them may be used to control the results.

Consider now the results of calculations shown in Table 3. Option 1 corresponds to energy explosion of 200 megatons from Wikipedia. It can be seen that there may not be tree-felling in jungle on either coasts of the Sunda Strait at this explosion energy. A distance, where the shock wave may reach the desired intensity, is lesser at ~ 7.5 km than the length between the epicenter and the nearest point of north-eastern cape of Sumatra. And there couldn't be a significant amount of shattered windows in Batavia, not to mention about knocking-out of doors.

In the option 2 with 400 megatons of explosion energy is possible sevenfold fixing of infrasound wave from the blast with the aid of a sensitive barograph that corresponds to the minimum number of registrations described in the sources [27, 28]. However, this explosion energy is not enough to ensure that in the vicinity of Krakatoa could be observed that was in August 1883. A shock wave with the required overpressure for tree-felling only can touch the nearest beach in the north-east. And such amount of shattered glass windows in Batavia was practically impossible.

Overpressure peak reaches 5 kPa at the distance of 155 km, and this overpressure below 7 kPa value of less than 10 % at the distance of 130 km, in the option 3 with 1040 megatons explosion energy. The distance of tree-felling is only less on 700 m in this case than earlier predetermined distance 50 km. So, taking into account that the nominal boundary conditions are approximate, we may assume that the option 3 satisfies by them, and this explosion energy is close to the minimum value, which may be matched with observations.

Option 4 with 1090 megaton explosion energy satisfies to the tree-felling boundary condition, and perfectly consistent with what happened in Batavia. A version 5 with 1200 megaton of explosion energy looks already in general a bit excessive, especially if we take into account excess of the path length of its infrasound wave. Thus, we can conclude that the energy of the Krakatoa explosion was close to the value of $E_e$ = 1.09 ± 0.05 Gt, which is almost 50 times higher than the minimum and is 10 times lower than its maximum in prior estimations. With the exception of extremes, the resulting value is about 5 times more than the minimum and 5 times less than the maximum data. That is it appeared practically in the center of all this field of scattering of earlier made assumptions about the energy of Krakatoa explosion. It should be noted that a similar approach as was described here for evaluating of Tambora explosion energy leads to a value $E_e$ = 5.4 ± 0.3 Gt that 4.9 times greater than in case of Krakatoa blast. It is virtually the same as the conventional estimates of the ratio of obtained earlier for Tambora and Krakatoa explosions energies, probably through the volume of discarded the eruptive products – about 100 km$^3$ and 20 km$^3$, respectively [27, 28, 40, 41, 44].

It is worth mentioning here about Rayabasa – volcano, which is standing on the coast of Sumatra Island, 42 km from Krakatoa [47] (label 2 in Fig. 2 indicates the location of this mountain). It was inside of the nominal tree-felling zone close to its border. This mountain has masked a region, lying on the opposite side from Krakatoa, from the shock wave, creating so-called «wind shadow», and dramatically reducing the impact of shock wave out there on the terrain. At this distance from the epicenter of the explosion overpressure in the wave on the open countryside would be for about 41 kPa (at $E_e$ = 1.09 Gt). Even diminishing it to 1.5 times would lead to the cessation of tree-felling. So the jungle should to survive in the area of the «shadow».

And it is valid, still, according to description of one Indonesian tourist who has visited atop of the mountain of Rayabasa about 4 years ago. The beach from there is well visible because of absence of trees, and back slopes of Rayabasa opposite the sea (and Krakatoa) were overgrown densely by trees [48]. So even a century and a quarter after the disaster is clearly seen that the boundary of tree-felling was moved here closer to the epicenter at 8 km, due to the existence of the huge obstacle – the mountain, which height is of almost 1.3 km. In this zone, the shock wave has not destroyed even forest hut of Bering's family, which is known from history of the disaster [43], in spite of that the total destruction of wooden houses occurs usually at the overpressure peak on the shock wave from 20 to 30 kPa [32]. And with this sharp reduction of the tree-felling zone in the closest to the epicenter, it would be difficult to speak about shock wave, «which brought down the equatorial forests on Sunda Strait coast», at energies of Krakatoa explosion, which are lower than shown here.

**Conclusions**

1. Two independent acoustic methods have validated the estimation of earlier calculated value of explosion energy for Chelyabinsk meteoroid of about 57 Mt.
2. The first of these methods, as well as the evaluations of the energy through seismic signals and barograms, confirmed the explosion energy of Tunguska meteoroid at the level of 14.0 – 14.5 Mt.
3. We also obtained an excellent agreement between the acoustic estimates and other data on explosion energy of another meteoroid and two catastrophic volcanoes blasts – Bezymyanny and Krakatoa.

**References**

1. Yu. Lobanovsky – Parameters of Chelyabinsk and Tunguska objects and features of explosions caused by them. *Synerjetics Group*, April 12, 2013 // http://www.synerjetics.ru/article/objects.htm (in Russian)
2. Yu. I. Lobanovsky – Parameters of Chelyabinsk and Tunguska objects and their explosion modes. *Arxiv.org*, July 07, 2013 // http://arxiv.org/abs/1307.1967
3. Yu. I. Lobanovsky – Adequacy evaluation of the conditions on the shock wave far away from the explosion epicenter. *Synerjetics Group*, April 19, 2013 // http://synerjetics.ru/article/border.htm (in Russian)
4. Yu. Lobanovsky – Two families of comet fragments and a little about their parents. *Synerjetics Group*, April 24, 2013// http://www.synerjetics.ru/article/families.htm (in Russian)
5. Yu. Lobanovsky – Refined parameters of Chelyabinsk and Tunguska objects and features of their explosions. Synerjetics Group, January 26, 2014// http://www.synerjetics.ru/article/objects_2.htm (in Russian)
6. T. Khan –Interview with Vladimir Puchkov. *Rescuer. Emergency Department*, February 25, 2013 // http://spasatel-mchs.ru/edition/51432/document836703/ (in Russian)
7. A. Le Pichon, L. Ceranna and others – 2013 Russian Fireball largest ever detected by CTBTO infrasound sensors. *Geophysical Research Letters DOI: 10.1002/grl.50619* // http://onlinelibrary.wiley.com/doi/10.1002/grl.50619/abstract
8. K. Allen – Chelyabinsk fireball: Canadian scientists still sizing it up. *Thestar.com, World*, 12.04.2013 // http://www.thestar.com/news/world/2013/04/12/chelyabinsk_fireball_canadian_scientists_still_sizing_it_up.html
9. O. Popova, I. Nemchinov – Bolides in the Earth atmosphere. *In Book: Catastrophic Events Caused by Cosmic Objects.* Springer, 2008 // https://thepiratebay.sx/torrent/6701391/
10. A. Y. Fernandez, O. C. Jalabert – Infrasonidos y ondas acústicas de gravedad. *Universidad Austral de Chile*, Valdivia, 2002 // http://cybertesis.uach.cl/tesis/uach/2002/bmfcip821i/doc/bmfcip821i.pdf (in Spanish)
11. Physics of explosion and shock, ed. by L. P. Orlenko, vol. 1. Moscow, Fizsmatgiz, 2002 // http://padabum.com/ (in Russian)
12. The distribution of the energy released by nuclear explosions. *Nuclear Weapons* // http://vault-13.ru/t2/c24/ (in Russian)
13. Tsar Bomba. *Wikipedia* // http://ru.wikipedia.org/wiki/Царь-бомба (in Russian)
14. 1976 Standard Atmosphere Calculator // http://www.digitaldutch.com/atmoscalc/
15. V. B. Adamski, Yu. N. Smirnov – 50-megaton explosion over Novaya Zemlya. *Soviet atomic project: People and Events* // http://www.wsyachina.narod.ru/history/50_mt_bomb.html (in Russian)
16. IMS Infrasound Network. *DTRA Verification Database* // http://www.rdss.info/infrastat/network/map.html
17. E. Farkas – Transit of Pressure Waves through New Zealand from the Soviet 50 Megaton Bomb Explosion. *Nature,* no. 4817, February 24, 1962
18. D. O. ReVelle – Historical detection of atmospheric impacts by large bolides using acoustic-gravity waves. *International Symposium on Near-Earth Objects, United Nations/Nations/Explorers Club*, New York City, April 24 – 26, 1995 // http://adsabs.harvard.edu/abs/1997NYASA.822..284R
19. New Releases. *US Air Force* // https://www.fas.org/irp/agency/aftac/news/index.html
20. L. A. Kulik – Data on Tunguska meteorite to 1939. *Reports of Academy of Sciences of USSR*, vol. XXII, no 8, 1939 // http://tunguska.tsc.ru/ru/science/bib/1930-39/1939/1939-01/?print=on (in Russian)
21. A. V. Zolotov – Problems of Tunguska catastrophe in 1908. Minsk, «Science and Technology», 1966 // http://tunguska.tsc.ru/ru/science/1/zol/5/24/?print=on (in Russian)
22. Bezymyanny (Volcano). *Wikipedia* // http://ru.wikipedia.org/wiki/Безымянный_(вулкан) (in Russian)
23. G. S. Gorshkov – Eruption of volcano Bezymyanny (Preliminary Report). *Academy of Sciences of the USSR, Bulletin of Volcanological Station,* no 26, 1956 // http://www.kscnet.ru/ivs/publication/art_bezym/1956.pdf (in Russian)

24. A. Ben-Menahem – Source parameters of the Siberian explosion of June 30, 1908, from analysis and synthesis of seismic signals at four stations. *Physics of the Earth and Planetary Interiors*, vol. 11, no. 1, 1975 // http://65.54.113.26/Publication/40382391/source-parameters-of-the-siberian-explosion-of-june-30-1908-from-analysis-and-synthesis-of
25. V. Svetsov, V. Shuvalov – Tunguska Catastrophe of 30 June 1908. *n Book: Catastrophic Events Caused by Cosmic Objects.* Springer, 2008 // http://confessedtravelholic.com/doc/download-ebook-Catastrophic-Events-Caused-by-Cosmic-Objects-pdf-amazon-304051.html
26. R. Marcus, H. J. Melosh, G. Collins – Earth Impact Effects Program. *Imperial College (London), Purdue University* // http://impact.ese.ic.ac.uk/cgi-bin/crater.cgi?dist=20&diam=17&pdens=1000&pdens_select=0&vel=18&theta=30&tdens=&tdens_select=3000
27. Krakatoa. *Wikipedia* // http://en.wikipedia.org/wiki/Krakatoa
28. Krakatoa. *Wikipedia* // http://ru.wikipedia.org/wiki/Кракатау (in Russian)
29. I. A. Rezanov – Great catastrophes in the history of the Earth. Moscow, Science, 1984 // http://www.twirpx.com/file/629113/ (in Russian)
30. IMS Infrasound Network. *Natural Resources Canada* // http://can-ndc.nrcan.gc.ca/is_infrasound-eng.php
31. P. G. Brown at al – A 500-kiloton airburst over Chelyabinsk and an enhanced hazard from small impactors. *Nature Latter*, **503**, no 7475, 14.11.2013 // http://www.nature.com/nature/journal/v503/n7475/full/nature12741.html
32. Atmospheric nuclear explosion. *Wikipedia* // http://ru.wikipedia.org/wiki/Атмосферный_ядерный_взрыв (in Russian)
33. Chronology of the USSR nuclear tests (1949 – 1962). *Wikipedia* // http://ru.wikipedia.org/wiki/Хронология_ядерных_испытаний_СССР_(1949—1962) (in Russian)
34. Ivy Mike. *Wikipedia* // http://en.wikipedia.org/wiki/Ivy_Mike
35. Castle Bravo. *Wikipedia* // http://en.wikipedia.org/wiki/Castle_Bravo
36. Operation Dominic 1962. *Janes's Oceania Home Page* // http://www.janeresture.com/christmas_bombs/operation_dominic_1962.htm
37. Earth science: An illustrated guide to science. Chelsea House Publishers, 2006 // http://geology.lnu.edu.ua/phis_geo/fourman/library-Earth/Earth%20Science%20An%20Illustrated%20Guide%20to%20Science.pdf
38. E. Jones, J. Kodis – Atmospheric effects of large body impacts: The first few minutes. *In: Geological implications of impacts of large asteroids and comets on the Earth.* The Geological Society of America, Special Paper 190, 1982 // http://books.google.ru/books?id=efz87W88HawC&pg=PA175&lpg=PA175&dq=krakatau+%221024+ergs%22&source=bl&ots=9KOfFcf1I0&sig=88SlLh0OuKoPPQ5r-HlGGIj1JTY&hl=en&sa=X&ei=MaI0UrXUD-fl4AP0goDYBQ&ved=0CD8Q6AEwAw#v=onepage&q=krakatau%20%221024%20ergs%22&f=false
39. Mount Tambora. *Wikipedia* // http://en.wikipedia.org/wiki/Mount_Tambora
40. V. I. Vaganov et al. – Blasting ring structures of shields and platforms. Moscow, Nedra 1985, (in Russian)
41. The Year without Summer. *A Cicada Film for Discovery Channel*, 2003 // http://www.downloadprovider.me/download-k:The-Year-Without-Summer.html?aff.id=1299
42. R. Hoblitt, C. Dan Miller, W. Scott – Volcanic hazards with regard to sitting nuclear-power plants in the Pacific Northwest. *US Department of the Interior Geological Survey*, Open-File Report 87-297. 1987 // http://pubs.usgs.gov/of/1987/0297/report.pdf
43. Ultimate Blast: Eruption at Krakatau. *Terra Nova Television, Discovery Channel* // http://www.youtube.com/watch?feature=player_embedded&v=E8Sg6LvyUPo#t=2878
44. A. V. Vikulin – World of vortex motions. Petropavlovsk-Kamchatsky, KamchatGTU, 2008 // http://www.kscnet.ru/ivs/monograph/vikulin_2/ (in Russian)
45. Dutch East India Company. *Wikipedia* // http://en.wikipedia.org/wiki/Dutch_East_India_Company
46. Jakarta. *Wikipedia* // http://en.wikipedia.org/wiki/Jakarta
47. Rajabasa. *Wikipedia* // http://en.wikipedia.org/wiki/Rajabasa
48. F. Seponada – Misteri gunung Rajabasa, 2009 // http://wisata.kompasiana.com/jalan-jalan/2009/12/29/misteri-gunung-rajabasa-45270.html (in Indonesian)